\documentclass[11pt]{article}
\usepackage{graphicx}

\newcommand{\BABARPubYear}    {05}

\newcommand{\BABARConfNumber} {15}
\newcommand{\SLACPubNumber} {11364}
\newcommand{\LANLNumber} {0507075}

\input pubboard/babarsym

\long\def\inst#1{\par\nobreak\kern 4pt\nobreak
    {\it #1}\par\vskip 10pt plus 3pt minus 3pt}
\def\myprl  #1 #2 #3 {\jprl{#1},\ #2 (#3)}
\def\myplb  #1 #2 #3 {\plb{#1},\ #2 (#3)}
\def\myprd  #1 #2 #3 {\jprd{#1},\ #2 (#3)}
\def\mynim  #1 #2 #3 {\nim{#1},\ #2 (#3)}
\def\mypr   #1 #2 #3 {\pr{#1},\ #2 (#3)}

\def\sss{\scriptscriptstyle}
\def\barpd{{\raise.35ex\hbox
{${\sss (}$}}--{\raise.35ex\hbox{${\sss )}$}}}
\def\dbarp{\hbox{$D^{0}$\kern-1.25em\raise1.5ex\hbox{\barpd}}}

\begin{document}
{\pagestyle{empty}

\begin{flushright}
\babar-CONF-\BABARPubYear/\BABARConfNumber \\
SLAC-PUB-\SLACPubNumber \\
hep-ex/\LANLNumber \\
July 2005 \\
\end{flushright}

\par\vskip 5cm

% Title of the paper
\begin{center}
\Large \bf Measurement of {\boldmath \CP}-violating parameters in fully reconstructed {\boldmath $B{\to}D^{(*)\pm}\pi^{\mp}$} and {\boldmath $B{\to}D^{\pm}\rho^{\mp}$} decays
\end{center}
\bigskip

\begin{center}
\large The \babar\ Collaboration\\
\mbox{ }\\
\today
\end{center}
\bigskip \bigskip

% Abstract
\begin{center}
\large \bf Abstract
\end{center}
We present a  preliminary measurement of  the \CP-violating parameters in 
fully reconstructed $\Bz{\to}D^{(*)\pm}\pi^{\mp}$ and $\Bz{\to}D^{\pm}\rho^{\mp}$ decays in 
 approximately $232$ million \Y4S $\to$ \BB decays collected with the
\babar\ detector at the PEP-II asymmetric-energy $B$ factory at SLAC.
From a  maximum likelihood fit to the time-dependent decay distributions we obtain for the parameters related to the \CP violation quantity $\sin(2\beta+\gamma)$:
\begin{eqnarray*}
\begin{array}{rclcrcl}
a^{D\pi}&=&-0.013 \pm 0.022 \ (\mbox{stat.}) \pm  0.007\ (\mbox{syst.})&\!\!\!\!,&
c_{lep}^{D\pi}&=&-0.043\pm 0.042 \ (\mbox{stat.}) \pm  0.011\ (\mbox{syst.})\,,\\ \nonumber
a^{D^*\pi}&=&-0.043 \pm 0.023 \ (\mbox{stat.}) \pm  0.010\ (\mbox{syst.})&\!\!\!\!,&
c_{lep}^{D^*\pi}&=&\phantom{-} 0.047 \pm 0.042 \ (\mbox{stat.}) \pm  0.015\ (\mbox{syst.})\,,\\ \nonumber
a^{D\rho}&=&-0.024 \pm 0.031 \ (\mbox{stat.}) \pm  0.010\ (\mbox{syst.})&\!\!\!\!,&
c_{lep}^{D\rho}&=&-0.098\pm 0.055 \ (\mbox{stat.}) \pm  0.019\ (\mbox{syst.})\,.\\ \nonumber
\end{array}
\end{eqnarray*}

\vfill
\begin{center}
Presented at the 
International Europhysics Conference On High-Energy Physics (HEP 2005),
7/21---7/27/2005, Lisbon, Portugal
\end{center}

\vspace{1.0cm}
\begin{center}
{\em Stanford Linear Accelerator Center, Stanford University, 
Stanford, CA 94309} \\ \vspace{0.1cm}\hrule\vspace{0.1cm}
Work supported in part by Department of Energy contract DE-AC03-76SF00515.
\end{center}

\newpage
} % end of pagestyle{empty}

% Input author list file (use the same list as for LP for the moment)
%\input pubboard/authors_jun2005.tex
\begin{center}
\small

The \babar\ Collaboration,
\bigskip

B.~Aubert,
R.~Barate,
D.~Boutigny,
F.~Couderc,
Y.~Karyotakis,
J.~P.~Lees,
V.~Poireau,
V.~Tisserand,
A.~Zghiche
\inst{Laboratoire de Physique des Particules, F-74941 Annecy-le-Vieux, France }
E.~Grauges
\inst{IFAE, Universitat Autonoma de Barcelona, E-08193 Bellaterra, Barcelona, Spain }
% ...

\end{center}\newpage

% The body of the paper starts here
\section{INTRODUCTION}
\label{sec:Introduction}
While the measurement of $\sin 2\beta$ is now a precision measurement~\cite{babar_sin2b,belle_sin2b},
the constraints on the other two angles of the Unitarity Triangle~\cite{CKM}, $\alpha$ and $\gamma$, are
still limited by statistics and by theoretical uncertainties.
We report on the measurement of \CP-violating asymmetries
in  $\Bz{\to} D^{(*)\pm} \pi^{\mp}$ and  $\Bz{\to} D^{\pm} \rho^{\mp}$
decays in $\FourS\to \BB$ decays, 
which are related to $\sin(2\beta+\gamma)$~\cite{sin2bg,fleischer}.
This analysis updates results already presented in Ref.~\cite{olds2bg} using a 
larger data sample ($232$ million instead of $110$ million \Y4S $\to$ \BB decays), and complements
the recently published
\babar\  measurement of \CP violation in $B \rightarrow D^{*\pm}\pi^{\mp}$ decays with a partial reconstruction technique~\cite{partial}.

The time evolution of $B^0{\to} D^{(*)\pm} h^{\mp}$ decays, where $h^{\mp}$ is a charged meson
made of $u$ and $d$ quarks, is
sensitive to $\gamma$ because the CKM-favored
decay amplitude $\Bzb{\to} D^{(*)+} h^-$, which is proportional to the CKM matrix
elements $V^{}_{cb}V^*_{ud}$, and the doubly-CKM-suppressed
decay amplitude $\Bz{\to} D^{(*)+}h^-$, which is proportional to $V_{cd}V^*_{ub}$, interfere due  to \BzBzb mixing. % (Fig.~\ref{fig:feyn}).
The weak phase difference between the two decay amplitudes is $\gamma$. When combined with $\BzBzb$ mixing, the
 total weak phase difference between the interfering amplitudes is $2\beta+\gamma$.
                                                                                                         
The decay rate distribution for $B^0{\to} D^{(*)\pm}h^{\mp}$ decays, neglecting the decay width difference, is given by
\begin{eqnarray}
f^{\pm}(\eta,\deltat) &=& \frac{e^{-\left|\deltat\right|/\tau}}{4\tau} \times [1 \mp S_\zeta \sin(\deltamd\deltat) \mp\eta C \cos(\deltamd\deltat)]\,,
\label{eq:fplus}
\end{eqnarray}
where $\tau$ is the \Bz lifetime, $\deltamd$ is the $\Bz\Bzb$ mixing frequency,
and $\deltat = t_{\rm rec} - \t_{\rm tag}$ is the time of the
$B^0\to D^{(*)\pm} h^{\mp}$ decay ($B_{\rm rec}$) relative to the decay of the other $B$ ($B_{\rm tag}$).
In this equation the upper (lower) sign refers to the flavor of $B_{\rm tag}$ as \Bz(\Bzb),
while $\eta=+1$ ($-1$) and $\zeta=+$ ($-$) for the final state $D^{(*)-}h^{+}$ ($D^{(*)+}h^{-}$).
In the Standard Model, the $S$ and $C$ parameters can be expressed as
\begin{eqnarray}
S_\pm = -\frac{2\textrm{Im}(\lambda_\pm)}{1+|\lambda_\pm|^2}\,, \hspace{0.4cm} {\rm and}
\hspace{0.4cm}
C=\frac{1-r^2}{1+r^2}\,,
\label{eq:cands}
\end{eqnarray}
where  $r\equiv |\lambda_+| = 1/|\lambda_-|$ and
\begin{eqnarray}
\label{eq:lambda}
 \lambda_\pm =\frac{q}{p}
\frac{A(\Bzb{\to} D^{\mp}h^\pm)}{A(\Bz{\to}D^{\mp}h^\pm)}
=r^{\pm 1}e^{-i(2\beta+\gamma\mp\delta)}.
\end{eqnarray}
 Here
$\frac{q}{p}$ is a function of the elements of the mixing
Hamiltonian~\cite{PDG}, and $\delta$ is the relative strong phase
between the
two contributing amplitudes.
In these equations, the parameters $r$ and $\delta$ depend on the choice of
the final state and will be indicated as  $r^{D\pi}$, $\delta^{D\pi}$, in the  $\Bz{\to}D^{\pm}\pi^{\mp}$ case, and $r^{D\rho}$, $\delta^{D\rho}$, in the  $\Bz{\to}D^{\pm}\rho^{\mp}$ case.
 For the $\Bz{\to}D^{*\pm}\pi^{\mp}$ mode the expression is similar with $r^{D^*\pi}$ and $\delta^{D^*\pi}$ (according to Ref.~\cite{fleischer} the strong phase is $\delta^{D^*\pi}+\pi$ in this mode, but this does not affect this measurement).                        
Interpreting the  $S$ and $C$ parameters
 in terms of the angles of the Unitarity Triangle requires knowledge of the $r$ parameters. 
%
%Since the amplitude in the numerator of Eq.~\ref{eq:lambda} is  suppressed
%with respect to that in the denominator, 
The $r$ parameters are expected to be small ($\sim 0.02$) and therefore cannot be extracted from the measurement
of $C$ with the current statistics. 
They can be estimated, assuming $SU(3)$ symmetry and neglecting contributions from annihilation diagrams,
 from the ratios of branching fractions  $\BR(\Bz{\to} D_s^{(*)+}\pi^-)/{\BR(\Bz{\to}
  D^{(*)-}\pi^+)}$ and $\BR(\Bz{\to} D_s^{+}\rho^-)/{\BR(\Bz{\to}
  D^{-}\rho^+)}$~\cite{sin2bg,olds2bg,partial}. Note that 
the $\Bz{\to} D_s^{*+}\pi^-$ and $\Bz{\to} D_s^{+}\rho^-$ decays~\cite{dsrho} have not yet been observed.

\section{THE \babar\ DETECTOR AND DATASET}
\label{sec:babar}
This measurement is based on $232$ million \Y4S $\to$ \BB decays
collected with the \babar\ detector at the \pep2\ storage ring.
The \babar\ detector is described elsewhere~\cite{detector}.
We use Monte Carlo simulation of the \babar\ detector based on
GEANT4~\cite{geant} to validate the analysis procedure and to estimate
some of the backgrounds.

\section{ANALYSIS METHOD}
\label{sec:Analysis}
The analysis strategy is similar to our previous publications on this 
topic~\cite{olds2bg,oldolds2bg}
and to other time-dependent measurements performed at \babar~\cite{babar_sin2b}.

We reconstruct the $B_{\rm rec}$ in the $D^{\pm}\pi^{\mp}$, $D^{*\pm}\pi^{\mp}$ and $D^{\pm}\rho^{\mp}$ final states.
Candidate $B \to D^{*\pm}\pi^{\mp}$ 
 decays are reconstructed with the $D^{*\pm}$ decaying to 
\dbarp$\ \pi^{\pm}$, where the \dbarp\ subsequently decays to one of the four modes
$K^{\mp}\pi^{\pm}$, $K^{\mp}\pi^{\pm}\piz$, $K^{\mp}\pi^{\pm}\pi^{\mp}\pi^{\pm}$, or $\KS\pi^{\pm}\pi^{\mp}$. Candidate $B \to D^{\pm}\pi^{\mp}$ and $B \to D^{\pm}\rho^{\mp}$ decays are reconstructed with
the $D^{\pm}$ decaying into $K^{\mp}\pi^{\pm}\pi^{\pm}$ and $\KS\pi^{\pm}$, and the $\rho^{\pm}$ decaying into $\pi^{\pm}\piz$.
The $D^{(*)\pm}$ candidates are then combined with either a single track or 
a track and a \piz candidate with invariant mass in the $\rho^{\pm}$ window,
$620< m(\pi^{\pm}\pi^0)<920$ $\rm{MeV}/\rm{c}^2$. Exploiting the spin properties of
the decay of a pseudo-scalar meson into a pseudo-scalar and a vector meson, we require  the 
cosine of the helicity angle $\theta_{hel}$, defined as the angle between 
the charged pion and the $D$ momentum in the $\rho^{\pm}$ rest frame,
 to satisfy $|\rm{cos}\theta_{hel}|>0.4$. No requirement is applied on any helicity angle 
of the $B \to D^{*\pm}\pi^{\mp}$ decay channel.
The event selection and candidate reconstruction are described in more details in Refs.~\cite{olds2bg,oldolds2bg}.

To identify the flavor of the $B_{\rm tag}$, we use multivariate algorithms that identify signatures in the $B$ decay products that determine (``tag'') the flavor to be either a $B^0$ or a \Bzb. 
Primary leptons from semi-leptonic $B$ decays are selected from identified electrons and muons and from isolated energetic tracks. The charges of identified kaons and soft pions from $D^{*+}$ decays are also used to extract flavor information. These algorithms are combined, taking into account the correlations among different sources of flavor information, and provide an estimate of the mistag probability. The tagging procedure has been improved with respect to the procedure used in our previous analysis~\cite{olds2bg} with the addition of information from low momentum electrons, $\Lambda \rightarrow p\pi$ decays, and correlations among identified kaon candidates. Each event with mistag probability less than $45\%$ is assigned to one of six hierarchical, mutually exclusive tagging categories. 
The lepton tagging category contains events with an identified lepton, while other events are divided into categories based on the mistag probability. The effective efficiency of the tagging algorithm, defined as $Q = \Sigma_i \epsilon_i(1-2w_i)^2$, where $\epsilon_i$ and $w_i$ are the efficiency and the mistag probability for the tagging category $i$, respectively, improves by $5\%$ (relative) over the algorithm used in Ref.~\cite{olds2bg}.

The time difference $\deltat$ is calculated from the measured separation along the
beam collision axis $\deltaz$ between the $B_{\rm rec}$ and $B_{\rm tag}$ decay vertices. We determine the  $B_{\rm rec}$ vertex from its charged tracks. 
The  $B_{\rm tag}$ decay vertex is obtained by fitting tracks that do not belong
to $B_{\rm rec}$, 
imposing constraints on the $B_{\rm rec}$ momentum and the beam-spot location. 
The $\deltat$ resolution is approximately 1.1 ps.

\begin{figure}[!htb]
\begin{center}
\includegraphics[height=18cm]{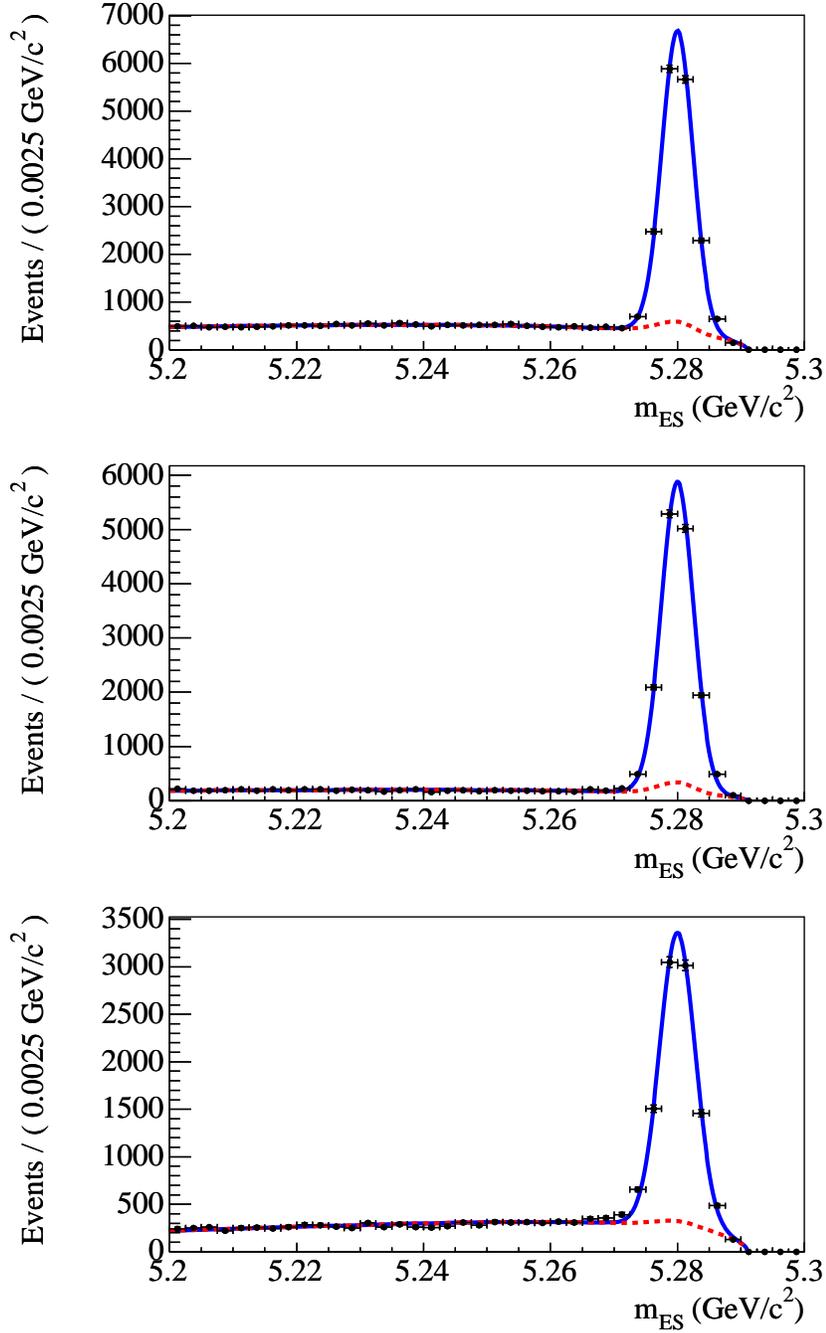}
\caption{\mes\ distributions in the signal region for, from top to bottom, the $B
  \rightarrow D^{\pm}\pi^{\mp}$, $B \rightarrow D^{*\pm}\pi^{\mp}$
and $B \rightarrow D^{\pm}\rho^{\mp}$ sample for events with tagging information. 
Each solid curve represents the result of a fit with a Gaussian (that describes signal events) 
plus a bifurcated Gaussian (that describes peaking background events) 
plus a threshold function (that describes combinatorial background events). 
Each dashed curve represents the background events only.
}
\label{mes}
\end{center}
\end{figure}

Signal and background are discriminated by two kinematic variables:
the beam-energy substituted mass, $\mes \equiv \sqrt{(\sqrt{s}/2)^{2} - {p_B^*}^2}$,
and the difference between the $B$ candidate's measured energy and the beam energy, 
$\DeltaE \equiv E_{B}^* - (\sqrt{s}/2)$.
$E_{B}^*$ ($p_B^*$) is the energy (momentum) of the \B\ candidate
in the $e^{+}e^{-}$ center-of-mass frame, and $\sqrt{s}$ is the total center-of-mass energy.
The signal region is defined to be $|\DeltaE| <3\sigma$, where the resolution $\sigma$ is mode-dependent
and is approximately 20\mev as determined from data. Figure~\ref{mes} shows the $\mes$ distribution for candidates in the signal region.

The background remaining after the event selection can be separated into two categories, one of which is due to random combinations
of particles in the event (${combinatorial }$ background) and the other is due to $B$ decays 
into final states similar to the signal final states, which have an \mes\ distribution similar to the signal
(${peaking}$ background). 

The $m_{ES}$ distribution of signal events is described by a Gaussian having a width of about 2.5\mevcc, whereas the peaking background is described by
a bifurcated Gaussian with widths of about 3\mevcc on the left side and 2.5\mevcc on the right side.
To separate the combinatorial background, the \mes\ distribution is fitted with the sum of a threshold function~\cite{argus} and the signal and peaking background components.

We estimate the contributions of background peaking in the \mes\ signal region using simulated events.
The most dominant peaking background modes are $B$ decays into open-charm final states similar to that of the
signal, $e.g.$ $B^0 \to D^{*-} K^+$, $B^+ \to \bar{D}^{*0}\pi^+/\rho^+$ or $\Bz\to D^{*-}\rho^+$.
The Gaussian yields and the amounts of peaking background are summarized in
Table ~\ref{tab:yields}. 
The latter are identified by their source as either coming from neutral or charged $B$ meson decays.
\begin{table}[ht]
\begin{center}
\caption{Signal yields, purities $P$, and fractions of peaking 
backgrounds $f_{peak}$ of the selected samples for events with
tagging information. The purity is computed in the signal region, for 
$m_{ES}$ in a three $\sigma$ region around the $B^0$ mass nominal value.\label{tab:yields}}
\begin{tabular}{|l|c|c|c|c|}\hline
Decay  & yields & $P(\%)$ & \multicolumn{2}{|c|}{$f_{peak}$(\%)} \\ 
mode &  & &  \Bz & \Bpm\\  
\hline
$B {\to}D^{\pm}\pi^{\mp}$    & 15635$ \pm $135   & 85.5$ \pm $0.3    & 3.3$ \pm $0.1    &  1.3$ \pm $0.1  \\
$B {\to}D^{*\pm}\pi^{\mp}$   & 14554$ \pm $126   & 93.0$ \pm $0.2    & 2.8$ \pm $0.1    &  0.8$ \pm $0.1 \\
$B {\to}D^{\pm}\rho^{\mp}$   & 8736$ \pm $105    & 81.7$ \pm $0.4   & 1.3$ \pm $0.2    &  1.5$\pm $0.2  \\
\hline
\end{tabular}
\end{center}
\end{table}
In the case of the $B^0 \rightarrow D^{\pm}\rho^{\mp}$ decays, an additional source of 
background must be considered, which has the same final state,
 $B^0 \rightarrow D^{\pm}\pi^{\mp}\piz$, where the $\pi^{\mp}\piz$ system is not
produced through the $\rho$ resonance. This background can introduce a  dependence of the $\lambda$ parameter of Eq.~\ref{eq:lambda} on the $\pi \pi^0$ invariant mass: its contribution has been studied in Ref.~\cite{olds2bg}. This part of the analysis is not updated in this paper; we use our old result in the evaluation of the systematic error (section~\ref{sec:Systematics}).

An unbinned maximum likelihood fit is performed to the time distribution of events in this sample. Taking into account 
 possible \CP violation effects on the tag side~\cite{DCSD} the likelihood for signal events of Eq.~1 
for each tagging category $i$ and for each decay mode $\mu=D^{\pm}\pi^\mp, D^{*\pm}\pi^\mp,
D^{\pm}\rho^\mp$ becomes:
\begin{eqnarray}
f^{\pm,{\mu}}_i(\eta,\deltat) &=&
  \frac{e^{-\left|\deltat\right|/\tau}}{4\tau} \times [ 1 \mp (a^{\mu}
  \mp \eta b_i - \eta c_i^{\mu}) \sin(\deltamd\deltat)\mp\eta\cos(\deltamd\deltat)]\,,
\label{completepdf}
\end{eqnarray}
where in the Standard Model
\begin{eqnarray}\nonumber
&a^{\mu}&=\ 2r^{\mu}\sin(2 \beta+\gamma)\cos\delta^{\mu}\,, \\ \nonumber
&b_i&=\ 2r^\prime_i\sin(2 \beta+\gamma)\cos\delta^\prime_i\,, \\
&c_i^{\mu}&=\ 2\cos(2 \beta+\gamma) (r^{\mu}\sin\delta^{\mu}-r^\prime_i\sin\delta^\prime_i)\,.
\label{acdep}
\end{eqnarray}
Here $r^\prime_i$ ($\delta^\prime_i$) is, for each tagging category, the effective amplitude (phase) 
used to parameterize the tag side interference.  
Terms of order $r^{\mu 2}$ and $r^{\prime 2}_i$ have been neglected.
Results are quoted only  for the six $a^\mu$ and $c^\mu_{lep}$ parameters,
 which are independent of the unknown parameters
$r^\prime_i$ and $\delta^\prime_i$ (semi-leptonic $B$ decays have no doubly 
CKM-suppressed  contributions and therefore $r^\prime_{lep}\equiv 0$).
The other parameters are constrained by 
the fit, but, as they depend 
on $r^\prime_i$ and
$\delta^\prime_i$, 
they do not contribute to the interpretation of the result in terms 
of $\sin(2\beta+\gamma)$.

The signal \deltat\ distribution
in Eq.~\ref{completepdf} is convoluted with the
resolution function parametrized with the sum of three Gaussians to take into account finite $\Delta t$ resolution.
The  probabilities of incorrect tagging ($w_i$) are accounted for by
multiplying the $a^{\mu}$, $c_i^{\mu}$
 parameters and the $\cos(\deltamd\deltat)$
term by the dilutions $D_i=1-2w_i$ (average dilutions for $B^0$ and \Bzb ). Possible differences in mistag probabilities for $B^0$ and \Bzb\ are also taken into account.
The resolution function and the tagging parameters 
are floated in the fit to the data and are consistent within errors with previous \babar\
 analyses~\cite{babar_sin2b}.

The likelihood function has a contribution for each source of background.
The combinatorial background is parametrized as the sum of a component
with zero lifetime and one with an effective lifetime floated in the fit to the data.
The fraction of each background component is determined from the events in the \mes\ sidebands, having $5.20<\mes<5.27\gevcc$.
The background events are described using effective dilution parameters, obtained from the fit to the data, and have a  
\deltat\ resolution function consisting of the sum of two Gaussians, also floated in the fit to the data. 
The charmed peaking background coming from $B^{\pm}$ mesons
is modeled by an exponential with the $B^{\pm}$ lifetime, and its amount is fixed to the value predicted by simulation. 
The resolution function is the same as the signal resolution, while the dilution parameters are fixed to the values obtained from a $B^+$ control sample.
The charmed and charmless peaking backgrounds from $B^{0}$
mesons, whose amounts are also fixed to the value estimated from simulation, 
are modeled with a likelihood similar to the signal likelihood with no \CP violation (all the $a$, $b$, $c$ parameters fixed to $0$). 
Possible \CP violation in this background is taken into account in the evaluation of the systematic uncertainties. 
The resolution and the dilution parameters are the same as for the signal.

\section{SYSTEMATIC STUDIES}
\label{sec:Systematics}
Table~\ref{tab:sys} shows the contributions to the systematic uncertainties of the $a$ and $c_{lep}$ \CP parameters. 
\begin{table}[ht]
\begin{center}
\caption{Systematic uncertainties on the $a$ and $c$ parameters.\label{tab:sys}}
\begin{tabular}{|l|l|l|l|l|l|l|}\hline
  & \multicolumn{2}{|c|}{ $\Bz{\to}D^{\pm}\pi^{\mp}$}& \multicolumn{2}{|c|}{$\Bz{\to}D^{*\pm}\pi^{\mp}$} & \multicolumn{2}{|c|}{$\Bz{\to}D^{\pm}\rho^{\mp}$}\\ \hline
Source & $\sigma_{a}$ & $\sigma_{c}$ & $\sigma_{a}$ & $\sigma_{c}$ & $\sigma_{a}$ & $\sigma_{c}$  \\
\hline
Vertexing ($\sigma_{\Delta t}$) & 0.0037& 0.0064& 0.0080 & 0.0110 & 0.0047& 0.0110 \\
Fit ($\sigma_{\rm fit}$) & 0.0051 & 0.0088 & 0.0052& 0.0093 & 0.0075& 0.0129 \\
Model ($\sigma_{\rm mod}$) & 0.0012 & 0.0013 & 0.0012& 0.0013 & 0.0001& 0.0018 \\
Tagging ($\sigma_{\rm tag}$) & 0.0007 & 0.0016 & 0.0011 & 0.0014&0.0006 
& 0.0012 \\
Background ($\sigma_{\rm bkg}$) & 0.0023 & 0.0029 & 0.0020& 0.0029 & 0.0042& 0.0070\\
Dependence on $m_{\pi\pi^0}$ ($\sigma_{\rm \lambda dep}$) & $-$ & $-$ & $-$ & $-$ & 0.0018 &0.0047 \\
\hline
Total ($\sigma_{\rm tot}$) & 0.0069& 0.0114&0.0099&0.0150 & 0.0100 &0.0193\\ \hline
\end{tabular}
\end{center}
\end{table}

The impact of a possible systematic mis-measurement of $\deltat$ ($\sigma_{\Delta t}$)
has been estimated by comparing  different
parameterizations of  the resolution function, varying
the position of the beam spot, and
 the absolute $z$ scale within their uncertainties, 
 and loosening and tightening the 
quality criteria on the reconstructed vertex. We also estimate the
impact of the uncertainties on the alignment of the silicon vertex tracker (SVT) by repeating the measurement using simulated events, intentionally
misaligning the SVT in the simulation.
As the systematic uncertainty of the fit ($\sigma_{\rm fit}$), 
 we quote the upper limit on the bias on the $a^{\mu}$ and $c^{\mu}$ parameters,  
 as estimated
from samples of fully-simulated events. The model error
($\sigma_{\rm mod}$) contains the uncertainty on the $B^{0}$ lifetime and \deltamd,
varied by the uncertainties on the world averages~\cite{PDG} and also by floating them in the fit. 
The tagging error ($\sigma_{\rm tag}$)
 is estimated
considering possible differences in tagging efficiency between \Bz\ and \Bzb\ and 
allowing for different $\Delta t$ resolutions for correctly and incorrectly tagged events. 
We also account for uncertainties in the background ($\sigma_{\rm bkg}$)
 by varying the effective lifetimes, dilutions, \mes~shape parameters, signal fractions, 
and background \CP asymmetry.
For the  $B\to D\rho$  decay we also include the maximum bias of the $a$ and $c_{lep}$ parameters due
 to the possible dependence of $\lambda$ on  the $\pi \pi^0$ invariant mass ($\sigma_{\rm \lambda dep}$), 
as discussed in section \ref{sec:Analysis}.

\section{PHYSICS RESULTS}
\label{sec:Physics}
\begin{figure}[!htb]
\begin{center}
\includegraphics[height=12cm]{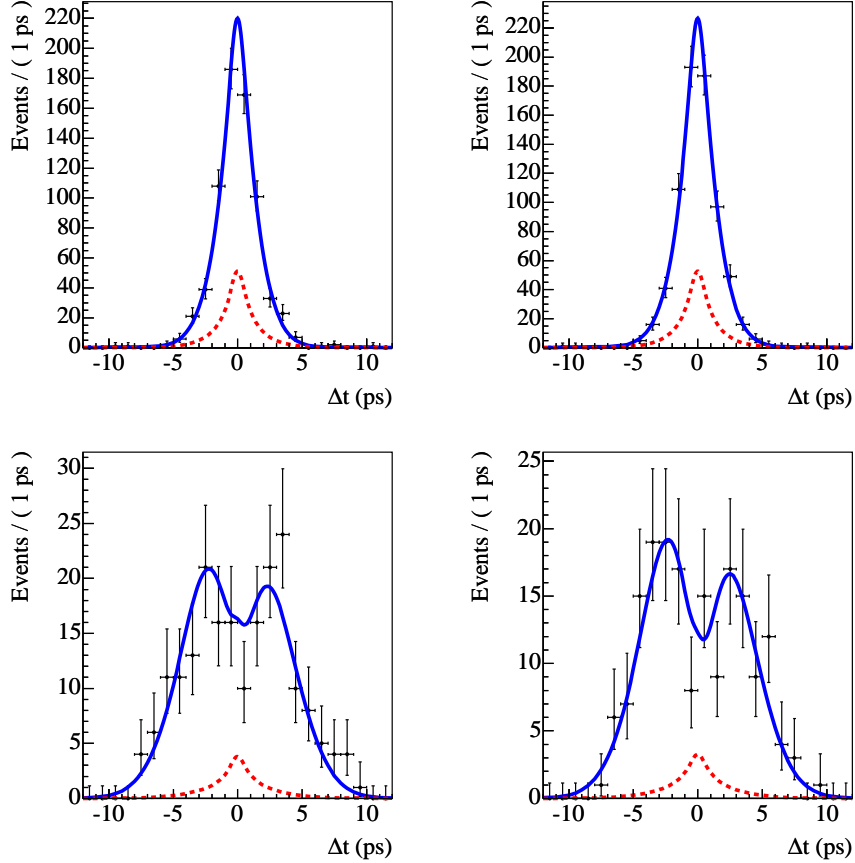}
\caption{$\Delta t$ distribution for data having the decay mode 
$B^0 \rightarrow D^-\pi^+$ in the lepton tagging category for, from upper left going clockwise, $B_{tag}=\Bzb$ and $B_{rec}=D^-\pi^+$, $B_{tag}=B^0$ and $B_{rec}=D^+\pi^-$, $B_{tag}=B^0$ and $B_{rec}=D^-\pi^+$, $B_{tag}=\Bzb$ and $B_{rec}=D^+\pi^-$. The result of the fit is superimposed. In each plot, the dashed curve represents the background contribution.}
\label{fig:dt1}
\end{center}
\end{figure}

\begin{figure}[!htb]
\begin{center}
\includegraphics[height=12cm]{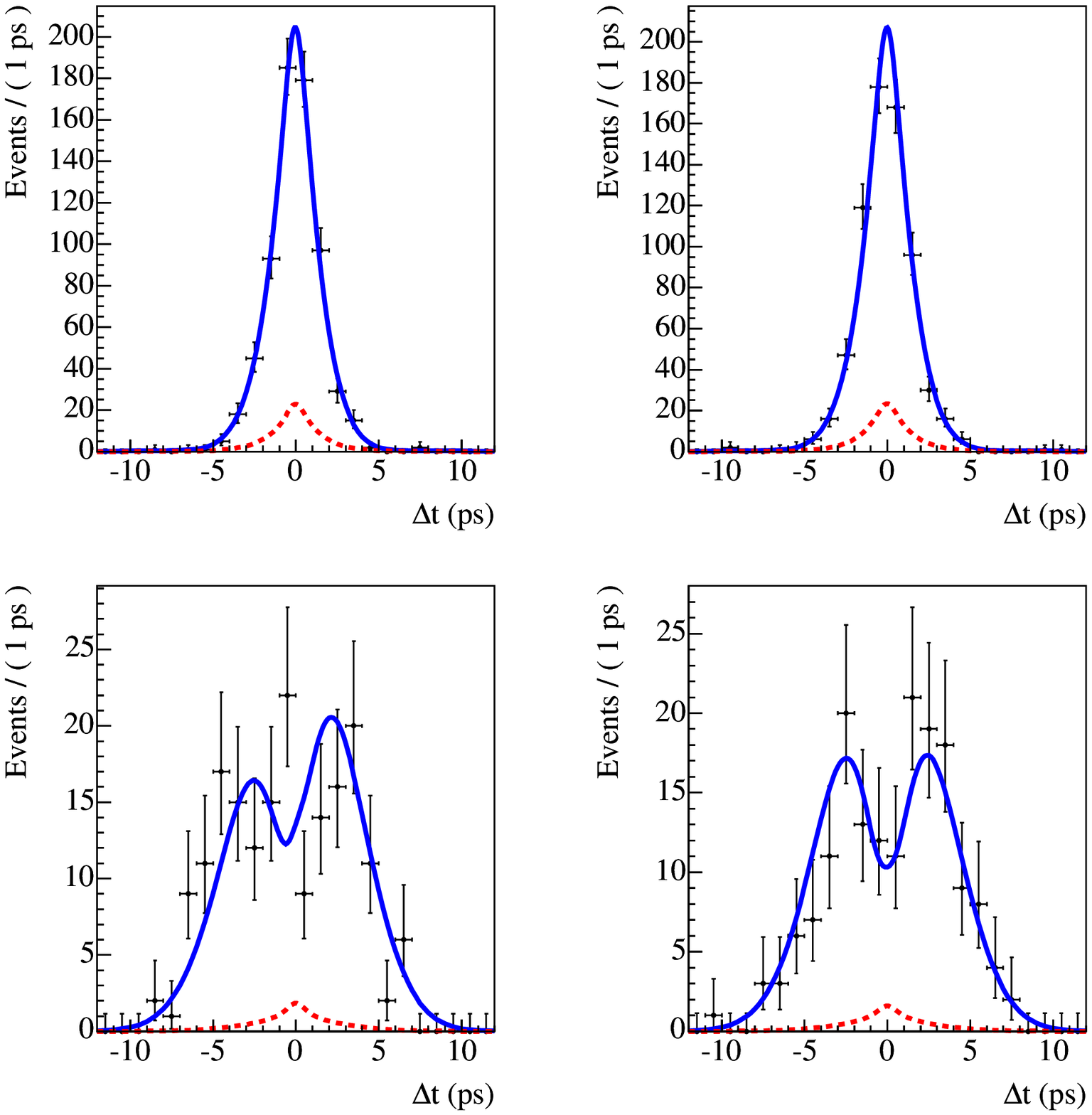}
\caption{$\Delta t$ distribution for data having the decay mode 
$B^0 \rightarrow D^{*-}\pi^+$ in the lepton tagging category for, from upper left going clockwise, $B_{tag}=\Bzb$ and $B_{rec}=D^{*-}\pi^+$, $B_{tag}=B^0$ and $B_{rec}=D^{*+}\pi^-$, $B_{tag}=B^0$ and $B_{rec}=D^{*-}\pi^+$, $B_{tag}=\Bzb$ and $B_{rec}=D^{*+}\pi^-$. The result of the fit is superimposed. In each plot, the dashed curve represents the background contribution.}
\label{fig:dt2}
\end{center}
\end{figure}

\begin{figure}[!htb]
\begin{center}
\includegraphics[height=12cm]{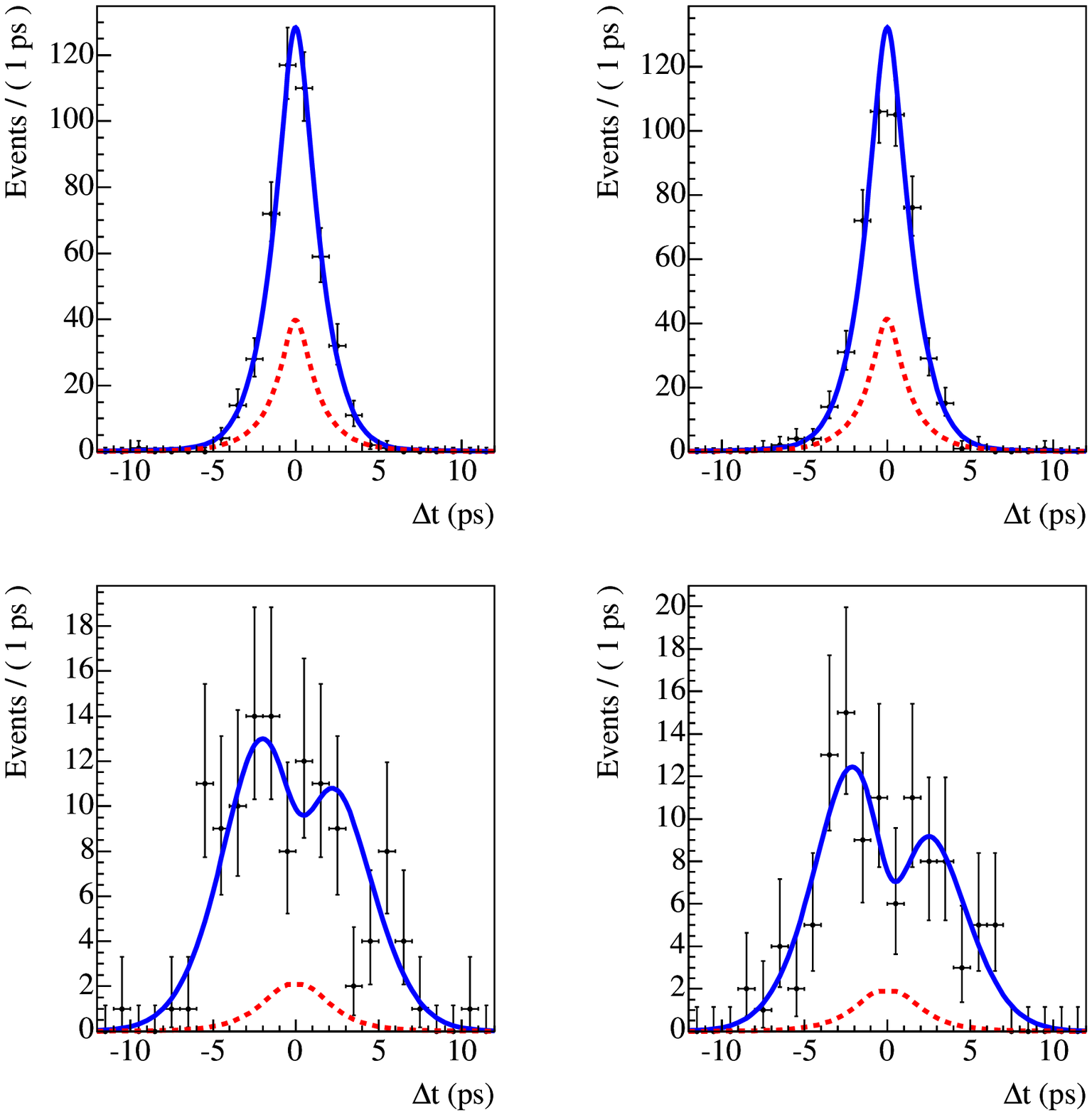}
\caption{$\Delta t$ distribution for data having the decay mode 
$B^0 \rightarrow D^-\rho^+$ in the lepton tagging category for, from upper left going clockwise, $B_{tag}=\Bzb$ and $B_{rec}=D^-\rho^+$, $B_{tag}=B^0$ and $B_{rec}=D^+\rho^-$, $B_{tag}=B^0$ and $B_{rec}=D^-\rho^+$, $\Bzb$ and $B_{rec}=D^+\rho^-$. The result of the fit is superimposed. In each plot, the dashed curve represents the background contribution.}
\label{fig:dt3}
\end{center}
\end{figure}

\begin{figure}[!htb]
\begin{center}
\includegraphics[height=10cm]{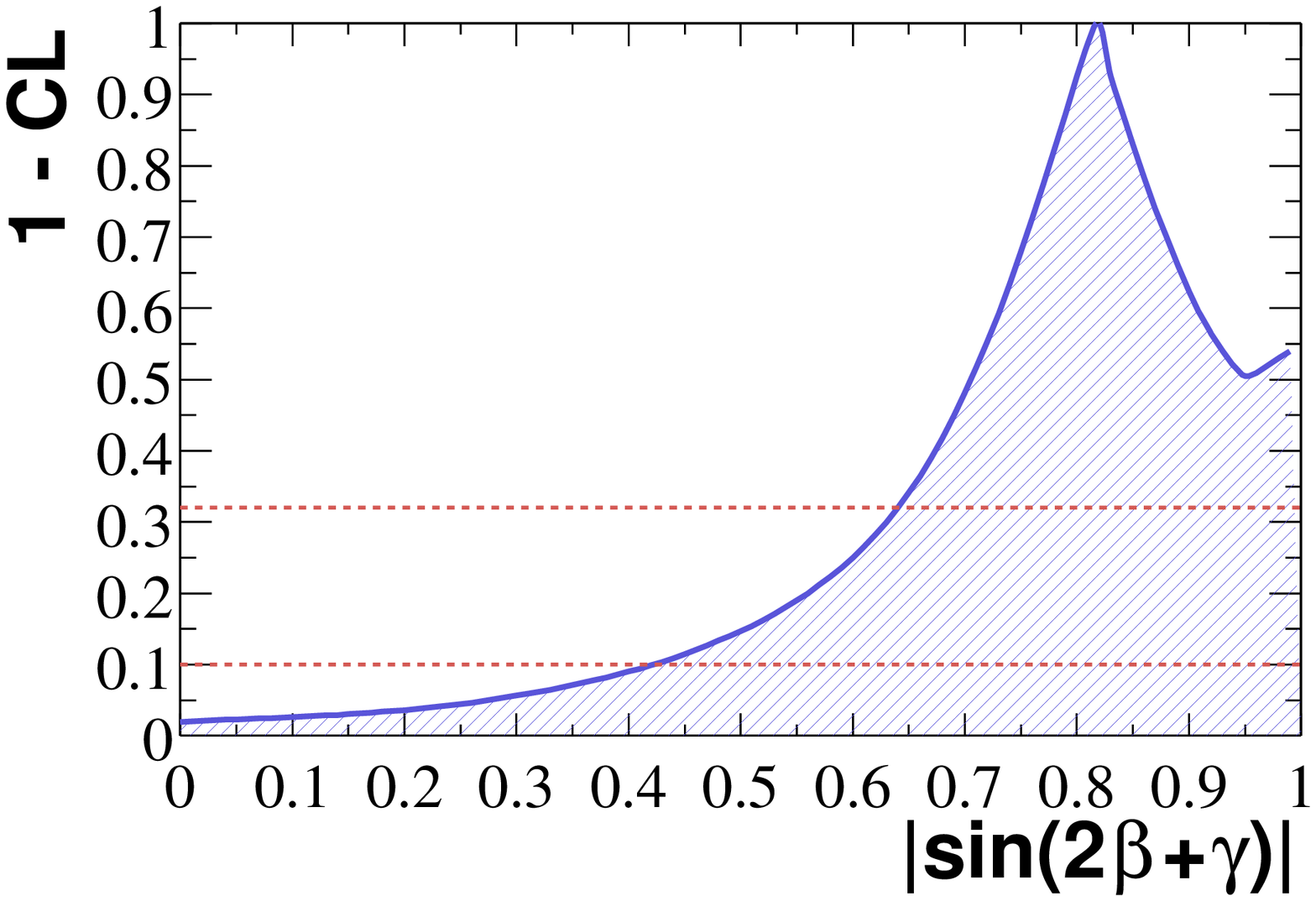}
\caption{Frequentistic confidence level as a function of $|\sin(2\beta + \gamma)|$,
obtained when combining our result with the result previously obtained on
partially reconstructed $B \rightarrow D^{*\pm}\pi^{\mp}$ decays~\cite{partial}.
The horizontal lines show, from top to bottom, the $68\%$ and $90\%$ confidence level.}
\label{fig:s2bg1}
\end{center}
\end{figure}

From the unbinned maximum likelihood fit we obtain the result:
\begin{eqnarray*}
\begin{array}{rclcrcl}
a^{D\pi}&=&-0.013 \pm 0.022 \ (\mbox{stat.}) \pm  0.007 \ (\mbox{syst.})&\!\!\!\!,&
c_{lep}^{D\pi}&=&-0.043\pm 0.042 \ (\mbox{stat.}) \pm  0.011 \ (\mbox{syst.})\,,\\ \nonumber
a^{D^*\pi}&=&-0.043 \pm 0.023 \ (\mbox{stat.}) \pm  0.010 \ (\mbox{syst.})&\!\!\!\!,&
c_{lep}^{D^*\pi}&=&\phantom{-}0.047 \pm 0.042 \ (\mbox{stat.}) \pm  0.015 \ (\mbox{syst.})\,,\\ \nonumber
a^{D\rho}&=&-0.024 \pm 0.031 \ (\mbox{stat.}) \pm  0.010 \ (\mbox{syst.})&\!\!\!\!,&
c_{lep}^{D\rho}&=&-0.098\pm 0.055 \ (\mbox{stat.}) \pm  0.019 \ (\mbox{syst.})\,.\\ \nonumber
\end{array}
\end{eqnarray*}
The largest correlation with any linear combination of other fit
parameters is about $20\%$ and $30\%$ for the $a^\mu$ and the $c_{lep}^\mu$ parameters respectively.
Figures~\ref{fig:dt1} to~\ref{fig:dt3} show the $\Delta t$ distribution for events tagged with leptons (which have the lowest mistag probability for the 
$B {\to}D^{\pm}\pi^{\mp}$ , $B {\to}D^{*\pm}\pi^{\mp}$ , $B {\to}D^{\pm}\rho^{\mp}$  modes).
We combine our result with the result obtained on partially reconstructed $B \rightarrow D^{*\pm}\pi^{\mp}$ decays~\cite{partial}
and we use the frequentistic method described in Ref.~\cite{partial} to set a constraint on $|\sin(2\beta+\gamma)|$. 
The confidence level as a function of $|\sin(2\beta +\gamma)|$ is shown in Fig.~\ref{fig:s2bg1}. We
set the lower limits $|\sin(2\beta +\gamma)|>0.64\ (0.42)$ $@$ $68\%\ (90\%)$ confidence level.
%Figure~\ref{fig:s2bg2} shows the corresponding probability contours for the apex of the Unitarity Triangle.
%
%
%

\section{SUMMARY}
\label{sec:Summary}

We have studied the time evolution of fully reconstructed
$\Bz{\to}D^{(*)\pm}\pi^{\mp}$ and $\Bz{\to}D^{\pm}\rho^{\mp}$ decays
in a data sample of $232$ million \Y4S $\to$ \BB decays.
\CP violation arising from the interference of the CKM-suppressed and the CKM-favored amplitudes 
 is expected to be small but sensitive to $\sin(2\beta+\gamma)$.

The measured \CP-violating parameters defined in Eq.~\ref{acdep} are 
shown in Section~\ref{sec:Physics}.
No significant \CP asymmetry is observed thus far. Using a frequentistic approach and
combining our result with the \babar\ result of Ref.~\cite{partial}, we set the limits: 
 $|\sin(2\beta +\gamma)|>0.64\ (0.42)$ $@$ $68\%\ (90\%)$ confidence level.

\section{ACKNOWLEDGMENTS}
\label{sec:Acknowledgments}

% Specific acknowledgments for this paper; remove if not needed.

% Standard acknowledgments paragraph; must always be included.
We are grateful for the 
extraordinary contributions of our \pep2\ colleagues in
achieving the excellent luminosity and machine conditions
that have made this work possible.
The success of this project also relies critically on the 
expertise and dedication of the computing organizations that 
support \babar.
The collaborating institutions wish to thank 
SLAC for its support and the kind hospitality extended to them. 
This work is supported by the
US Department of Energy
and National Science Foundation, the
Natural Sciences and Engineering Research Council (Canada),
Institute of High Energy Physics (China), the
Commissariat \`a l'Energie Atomique and
Institut National de Physique Nucl\'eaire et de Physique des Particules
(France), the
Bundesministerium f\"ur Bildung und Forschung and
Deutsche Forschungsgemeinschaft
(Germany), the
Istituto Nazionale di Fisica Nucleare (Italy),
the Foundation for Fundamental Research on Matter (The Netherlands),
the Research Council of Norway, the
Ministry of Science and Technology of the Russian Federation, and the
Particle Physics and Astronomy Research Council (United Kingdom). 
Individuals have received support from 
CONACyT (Mexico),
the A. P. Sloan Foundation, 
the Research Corporation,
and the Alexander von Humboldt Foundation.

\end{document}